\documentclass[12pt]{article}
\usepackage{amssymb}
\usepackage{graphicx}
\usepackage{amsmath}

\numberwithin{equation}{section}

\input{tcilatex}

\begin{document}

\title{\textbf{On the Mott formula for the a.c. conductivity and binary
correlators in the strong localization regime of disordered systems}}
\author{W. Kirsch$^{1)}$, O. Lenoble$^{2)}$, L. Pastur$^{3)}\thanks{%
Also Institute for Low Temperature Physics, Kharkov, Ukraine}$ \\
%EndAName
$^{1)}$ Ruhr University, Bochum, Germany, \\
$^{2)}$ Centre de Physique Th\'{e}orique, Marseille, France, \\
$^{3)}$ University Paris 7, France \\
}
\date{}
\maketitle

\begin{abstract}
We present a method that allows us to find the asymptotic form of various
characteristics of disordered systems in the strong localization regime,
i.e., when either the random potential is big or the energy is close to a
spectral edge. The method is based on the hypothesis that the relevant
realizations of the random potential in the strong localization regime have
the form of a collection of deep random wells that are uniformly and
chaotically distributed in space with a sufficiently small density. Assuming
this and using the density expansion, we show first that the density of
wells coincides in leading order with the density of states. Thus the
density of states is in fact the small parameter of the theory in the strong
localization regime. Then we derive the Mott formula for the low frequency
conductivity and the asymptotic formulas for certain two-point correlators
when the difference of the respective energies is small.
\end{abstract}

%\date{}

\hskip 0.4cm PACS numbers: 05.60.Gg, 72.15.Rn, 72.80.Ng

%\date{}

\section{Introduction}

\label{s:int}

It is widely accepted and proved rigorously in many cases that
elementary excitations in disordered media are localized if the
disorder is strong enough or/and the energy of the excitations is
close enough to the band edges. The idea dates back to the famous
paper \cite{An:58} by P. Anderson who emphasized, in particular,
the aspects related to the transition from localized to
delocalized states. The idea was further developed by N. Mott and
I. Lifshitz (see e.g. their review works \cite{Mo-Da:71,Li:65}).
In particular, it was I. Lifshitz who singled out the regime of
high disorder or low energy where the localization is most
pronounced. This regime is now known as the strong localization
regime. According to I. Lifshitz, in this case, the pertinent
realizations of the random potential have the form of a collection
of deep potential wells which are so rare and whose form is so
irregular that the quantum mechanical probability for tunnelling
through a macroscopic number of the localization wells vanishes.

The study of localization and relevant physical characteristics of
disordered systems can be reduced to the study of moments of the density
operator $\rho _{E}=\delta (E-H)$, where $H$ is the (one-body) Hamiltonian
of the system. By using the coordinate representation, we can write the $l$%
-th moment ($l$-th correlation function) as follows:
\begin{equation}
K_{l}(x_{1},...,x_{l};y_{1},...,y_{l};E_{1},...,E_{l})=\langle \rho
_{E_{1}}(x_{1},y_{1})...\rho _{E_{l}}(x_{l},y_{l})\rangle ,  \label{Kl}
\end{equation}%
where the $\langle ...\rangle $ denotes averaging with respect to the
disorder.

The simplest case of the correlation function (\ref{Kl}), corresponding to $%
l=1$, $x_{1}=y_{1}=x$:
\begin{equation}
\rho (E)=\langle \rho _{E}(x,x)\rangle  \label{dos}
\end{equation}%
i.e., to the average of the local density of states $\rho _{E}(x,x)$, is
known as the density of states (DOS) of the system.

I. Lifshitz suggested a non-perturbative method of computing the asymptotic
form of the DOS in the strong localization regime \cite{Li:63}. The above
description of typical realizations of the random potential is implemented
in this method by the assumption of independent quantization of a quantum
particle in each localization well (see \cite{Li:65,LGP,Ef-Sh:84}), thus the
complete localization of a particle in an exponential neighborhood of each
well. A rigorous proof of the complete and exponential localization in the
strong localization regime was given by J. Fr\"{o}hlich and T. Spencer \cite%
{Fr-Sp:83,Ma-Sc,Pa-Fi:92}.

In both these important results of the localization theory the
tunnelling between the localization wells plays no significant
role. In Lifshitz's argument, other wells are simply ignored. The
crucial ingredient in the rigorous proof of the complete
localization in the strong localization regime is a rather
sophisticated probabilistic extension of the
Kolmogorov-Arnold-Moser theory (known as the multi-scale analysis)
which allows one to verify that tunnelling between wells is
strongly suppressed, and therefore does not change qualitatively
the picture, suggested by the independent wells quantization
assumption.

The DOS determines equilibrium properties of a disordered system in the
one-body approximation, i.e., of the ideal Fermi gas in a random external
field. The study of kinetic properties of the gas and of interaction effects
requires the knowledge of higher moments (\ref{Kl}) of the density operator $%
\rho _{E}$, especially the second moment $K_{2}$. Important quantities that
can be expressed via $K_{2}$ are the density-density correlator and the
current-current correlator \cite{Go, LGP}. These correlators allow us to
answer relevant questions concerning the nature of localization and the
behavior of the conductivity and other physical characteristics.

The complete localization of states in a certain interval of
energies implies that the zero temperature d.c. conductivity
vanishes if the Fermi energy lies in this interval (see
\cite{Ai-Gr} for a proof and a discussion). On the other hand,
since the energies of localized states are dense, the
zero-temperature a.c. conductivity is expected to be non-zero for
any non-zero frequency $\nu $ of the external field. It was N.
Mott who first proposed \textquotedblright
resonant\textquotedblright\ tunnelling between pairs of wells as a
mechanism of the low frequency a.c. conductivity in localized
systems \cite{Mo-Da:71}. According to Mott, one can view those
states, resulting from independent quantization in each
localization well (localization center in Mott's terminology), as
a kind of \textquotedblright bare\textquotedblright\ states. They
decay exponentially in the distance from the corresponding
localization center. Two (several) bare states with widely spaced
centers but with sufficiently close energies can
\textquotedblright resonate\textquotedblright . This leads to the
two-center states (resp. multi center states), whose energies are
exponentially close in the separation between the centers. The
condition for a pair of wells to be in resonance determines the
distance between resonating wells, thereby determining the
characteristic value of the dipole moment of two bare states of
wells, and the square of the dipole moment is, in essence, the
conductivity according to the linear response theory (see formula (\ref{Ku2}%
) below). This observation leads to the following asymptotic expression for
the low frequency conductivity:
\begin{equation}
\sigma (\nu ,E_{F})=A\rho ^{2}(E_{F})\nu ^{2}\left( \log \frac{\nu _{0}}{\nu
}\right) ^{d+1}  \label{Mott}
\end{equation}%
in the case, where
\begin{equation}
T\ll \nu \ll E_{F}.  \label{cond}
\end{equation}%
Here $T$ is the temperature, $\nu $ the frequency of an alternating external
field, $E_{F}$ the Fermi energy (supposed to be in the localized spectrum). $%
A$ and $\nu _{0}$ are determined by the fundamental constants and by the
random potential.

Formula (\ref{Mott}) was discussed in many works (see e.g. \cite%
{Ef-Sh:84,Ef,Ha-Jo:91,Go,Go-Co,LGP,Mo-Da:71,We-Co} and Section 5). However,
a consistent "first principle" derivation of the formula\ is still not
available in a general multi-dimensional case. We mean a derivation based on
the Kubo formula (see formulas (\ref{Ku1})--(\ref{Ku2}) below), in which the
two-point correlation function is computed for a given random potential in
the asymptotic regime (\ref{cond}).

The fact that such a derivation is still missing encourages us to present in
this paper a heuristic method that allows us to obtain formula (\ref{Mott})
and some other two-point correlation functions (i.e., (\ref{Kl}) for $l=2$
and $|E_{1}-E_{2}|:=\nu \ll |E_{1}|,|E_{2}|$), and that, we believe,
clarifies Mott's initial arguments.

The method is based on the above hypothesis on the form of
pertinent realizations of the random potential as systems of deep
and rare localization wells. Viewing the density of wells as a
small parameter of the theory, we apply a version of the virial
expansion to compute the leading contribution to the moments
$K_{l}$ of (\ref{Kl}) for $l=1,2$. In particular, by applying this
procedure to the DOS, we find that its leading order is the
density of the localization wells. This shows that the small
parameter of the theory is the DOS itself, whose smallness is
known to be an important condition for localization. Furthermore,
we find that the leading order of the pair correlation functions,
the a.c. conductivity in particular, is determined by two-center
states, resulting from resonant tunnelling between a pair of
localization wells, in agreement with Mott's ideas. This leads to
formula (\ref{Mott}) and, therefore, supports the idea of pair
approximation in Mott's derivation of (\ref{Mott}). Among our
other
results, we mention high peaks of some pair correlation functions (see (\ref%
{C1}) and (\ref{C2}) below), appearing in a neighborhood of the origin and
on the \textquotedblright resonating\textquotedblright\ distance, determined
by the frequency of the external field. Analogous peaks were found before in
the one-dimensional case for strong localization \cite{Ha-Jo:91} as well as
in the weak localization regime \cite{Go-Co}. However, in these cases, the
peaks are of the order $\rho ^{2}(E_{F})$, while in the general $d$%
-dimensional case, the peaks are of the order $\rho ^{2}(E_{F})\left( \log
\nu _{0}/\nu \right) ^{d-1}$, i.e.,  much bigger in the regime (\ref{cond})
(see also \cite{We-Co} for a similar result).

The paper is organized as follows. In Section 2 we outline the method. In
Section 3 the Mott formula (\ref{Mott}) is derived. In Section 4 we derive
asymptotic formulas for binary correlators and in Section 5 we comment on
our results and on their relations to known results. % XXXXXXXXXXXXXXXXX

\section{ Method}

\label{s:meth}

\subsection{Effective potential}

\label{ss:efpot}

It was already mentioned in the introduction that extensive studies of the
strong localization regime show that the phenomenon is determined by
realizations of the random potential, containing deep and rare potential
wells. For a potential unbounded below (like the Poisson potential (\ref%
{Pois}) below) the large parameter of the theory is the absolute value of
the energy and/or the amplitude of the potential. These two cases of the
strong localization regime are manifestations of the simplest mechanism of
localization: capturing a quantum particle in strong and rare fluctuations
of a random potential\footnote{%
We mention another localization mechanism : enhanced backscattering. The
mechanism is responsible for localization at high energies in the
one-dimensional case, and for weak localization effects in arbitrary
dimensions.}.

In other words, for an overwhelming majority of eigenfunctions $\psi _{j}$,
corresponding to the strongly localized part of the spectrum, there exists a
point $\xi _{j}$, the center of the localization well, such that $\psi _{j}$
decays as $\exp \{-|x-\xi _{j}|/r_{j}\}$. Here $r_{j}$ is the localization
radius of $\psi _{j}$. The localization centers have to be uniformly and
chaotically distributed in space and the distances between them have to be
much bigger than the typical localization radii and than the radii of the
localization wells. Hence, one has to expect an effective \textquotedblleft
decoupling\textquotedblright\ between the localization wells.

One obtains a simple form of this picture of the strong
localization regime by replacing the random Schr\"{o}dinger
operator by the direct sum of operators, each of them defined in a
certain cell, containing a single localization well. This
procedure of independent quantization in isolated cells is
supported by and even instrumental in studies of the density of
states, the interband light absorption coefficient, and other
spectral and
physical characteristics of disordered systems (see e.g. \cite%
{Mo:93,Ef-Sh:84,Ki-Pa:90,LGP}), as well as of the probability distribution
of spacings between adjacent energy levels (see \cite{Mo:81,Mi:95}) in the
strong localization regime. However, the procedure is not appropriate in
studies of transport properties of disordered systems. This is why we
replace the procedure of independent quantization in isolated cells by the
less restrictive assumption, according to which %It is natural to call
%these fluctuations the localization wells.
%It is also clear that a
%property of the random potential that facilitate the localization is a
%sufficiently smooth probability distribution of its random parameters and a
%sufficiently strong decay of spatial correlations. The property makes more
%unlikely for different localization wells to be of the same shape, thereby
%suppress the tunnelling between different localization wells.
%All this makes natural the ansatz according to which
relevant properties of the strong localization regime can be described,
assuming that any shortly correlated and smoothly distributed random
potential can be replaced by an (effective) potential of the form:
\begin{equation}
V_{eff}(x)=\sum_{j}v_{j}(x-\xi _{j}).  \label{Veff}
\end{equation}%
Here $\{\xi _{j}\}$ are the Poisson random points of the density $\mu ,$
modelling the centers of the localization wells, and the random functions \{$%
v_{j}$\} are independent of each others and independent of the $\{\xi _{j}\}$%
. The \{$v_{j}$\} model the shape of the localization wells. We assume that
all $v_{j}$'s have a finite range and the typical radius $a$ of $v_{j}$'s is
related to the typical distance $\mu ^{-1/d}$ between wells as
\begin{equation}
a\ll \mu ^{-1/d}.  \label{amu}
\end{equation}%
The density $\mu $ of the localization centers is not known and has to be
found self-consistently. The density as well as the shapes of the wells may
depend on the energy interval in question.

In other words, we believe that the strong localization regime
possesses a certain robustness (insensitivity) with respect to a
concrete form of random potential, provided that it is translation
invariant in the mean, shortly correlated, and smoothly
distributed (the last two properties facilitate the localization
because they make it more unlikely that different localization
wells are of the same shape, thereby suppressing tunnelling
between different localization wells). One may say that our ansatz
(\ref{Veff}) replaces impenetrable walls between cells of the
independent quantization procedure by a kind of \textquotedblleft
soft\textquotedblright\ walls, that strongly suppress particle
mobility but do not exclude it completely.

To avoid technicalities,
%Therefore the relevant statistic of any random potential with the above
%properties can be described by the effective potential (\ref{Veff}) in the
%strong localization regime. Moreover,
we will choose a simple form of the localization wells, setting
\begin{equation}
v_{j}(x)=g_{j}v\left( \sqrt{g_{j}}x\right) ,  \label{vj}
\end{equation}%
where $v(x)$ is a finite range potential well and \{$g_{j}$\} are
independent identically distributed random variables, independent of \{$\xi
_{j}$\} and assuming arbitrary big positive values according to a smooth
probability density $p(g)$.

Summarizing, we can write the following formula for the effective potential
\begin{equation}
V_{eff}(x)=\sum_{j}g_{j}v\left( \sqrt{g_{j}}(x-\xi _{j})\right) .
\label{Veff1}
\end{equation}%
It should be noted that similar random functions are widely used in
localization theory as \textquotedblright bare\textquotedblright\ random
potentials in the Schr\"{o}dinger equation (see e.g. \cite{Ki,LGP}). We mean
the potentials of the form
\begin{equation}
V(x)=\sum_{j}\theta _{j}u(x-x_{j}),  \label{pot}
\end{equation}%
where $u$ is a non-positive function of a finite range (the single-impurity
potential). In the case, where $\{\theta _{j}\}$ are independent identically
distributed random variables and $\{x_{j}\}$ form a regular lattice, the
potential models a substitutional alloy, and in the case, where $\theta
_{j}=\theta =\mathrm{const}$ for all $j$ and $\{x_{j}\}$ are completely
chaotic (Poisson) random points of the density $c$, the potential
\begin{equation}
V(x)=\sum_{j}\theta \, u(x-x_{j}),  \label{Pois}
\end{equation}%
models an amorphous medium. Assuming that $c$ is large, $\theta $ is small
but $c\theta ^{2}=D$ is fixed and shifting the energy by the mean value
\begin{equation*}
c\theta \int u(y)dy
\end{equation*}%
of the potential (\ref{Pois}), we obtain a Gaussian random potential with
zero mean and with the correlation function
\begin{equation*}
D\int u(x-y)u(y)dy.
\end{equation*}%
In a more general case, where the $\{x_{j}\}$ are completely chaotic and the
$\{\theta _{j}\}$ are identically distributed random variables, independent
of each others and of the $\{x_{j}\}$, (\ref{pot}) is a generalized Poisson
potential.

We would like to stress here that while our effective potential (\ref{Veff1}%
) is similar to a generalized Poisson one (because of random
$g_{j}$'s), these two should not be identified. In particular, the
density $c$ of the impurity centers $\{x_{j}\}$ in (\ref{pot}) is
not the density $\mu $ of the localization centers $\{\xi _{j}\}$
in (\ref{Veff1}) ($\mu $ is usually is much smaller than $c$), and
the functions $\theta _{j}u$ in (\ref{pot}), modelling the single
impurity potential, have little in common with the functions
$v_{j}$ in (\ref{Veff1}), modelling the form of the localization
wells. The latter are formed by sufficiently large and dense
clusters of impurities in which the inter-impurity distances are
much smaller than the typical distance $c^{-1/d}$ between
impurities centers $\{x_{j}\}$. For example, if the
\textquotedblleft bare\textquotedblright\ random potential is
given by (\ref{Pois}), then it can be shown than the number of
$x_{j}$'s in a typical localization well is of the order $\log
E/u(0)>>1$ \cite{LGP}.

\subsection{Density expansion}

\label{ss:dexp}

Recall that an important property of the effective potential is the small
density $\mu $ of the localization centers (cf (\ref{amu})). We describe now
a technique that will allow us to use this property.

Let $\{F_{l}(x_{1},...,x_{l})\}_{l\geq 0}$ be a system of functions of $l$ $%
d $-dimensional variables $x_{1},...,x_{l}$ ($F_{0}$ is a constant). We
denote the set $(x_{1},...,x_{l})$ as $X$. Suppose that the system $%
\{F_{l}\}_{l\geq 0}$ satisfies the following conditions (we do not indicate
explicitly the index $l$).

\bigskip

\noindent (i) Translation invariance: for any $d$-dimensional vector $a$
\begin{equation*}
F(X)=F(X+a),\ \text{where }X+a=(x_{1}+a,...,x_{l}+a).
\end{equation*}

\noindent (ii) Additive clustering:
\begin{equation}
F(X\cup (Y+a))-\left[ F(X)+F(Y)\right] \rightarrow 0,\ \text{as }%
a\rightarrow \infty .  \label{ac}
\end{equation}
and the decay of the l.h.s. of (\ref{ac}) is fast enough (it will be
exponential below).

\bigskip

For any system of functions, possessing these properties, we can write the
combinatorial identity
\begin{equation}
F(X)=\sum_{Y\subset X}\sum_{Z\subset Y}(-1)^{N(Y\setminus Z)}F(Z),
\label{comb}
\end{equation}
where $N(X)$ is the number of points of $X$.

We will use this identity in the case, where $X,Y,Z$ are the sets of random
Poisson points $\{\xi _{j}\}$, entering in the effective potential (\ref%
{Veff}). Recall that an infinite system $\{\xi _{j}\}$ of Poisson points of
density $\mu $ in the $d$-dimensional space can be asymptotically described
as a system of random points $\xi _{1},...,\xi _{N}$, uniformly distributed
in a cube $\Lambda $, provided that the \textquotedblright
thermodynamic\textquotedblright\ limit $N\rightarrow \infty $, $|\Lambda
|\rightarrow \infty $ and $N/|\Lambda |\rightarrow \mu $ is carried out (we
will denote this limiting transition by $\Lambda \rightarrow \infty $). By
using this fact and identity (\ref{comb}), we can write that
\begin{eqnarray}
\lim_{\Lambda \rightarrow \infty }|\Lambda |^{-1}\langle F_{N}(\xi
_{1},...,\xi _{N})-F_{0}\rangle &=&\mu (F_{1}-F_{0})  \label{exp1} \\
&+&\frac{\mu ^{2}}{2}\int [F_{2}(x)-2F_{1}+F_{0}]dx+...  \notag
\end{eqnarray}%
where the symbol $\langle ...\rangle $ in the l.h.s. denotes averaging with
respect to the Poisson points $\{\xi _{j}\}$.

In view of (\ref{Veff1}), we will need a more general formula in which the
role of $\xi _{j}$'s is played by pairs $(\xi _{j},g_{j})$, where $\{g_{j}\}$
is a system of independent random variables of common density $p(g)$ which
are also independent of the $\{\xi _{j}\}$. The corresponding formula can be
obtained from (\ref{exp1}), written for fixed $g_{j}$'s and subsequently
integrated with respect to $g_{j}$'s with the probability density $p(g)$.
This yields
\begin{align}
\lim_{\Lambda \rightarrow \infty }|\Lambda |^{-1}\langle F_{N}((\xi
_{1},g_{1}),& ...,(\xi _{N},g_{N}))-F_{0}\rangle =\int
(F_{1}(g_{1})-F_{0})\mu (g_{1})dg_{1}  \label{exp2} \\
& +\frac{1}{2}\int [F_{2}(x;g_{1},g_{2})-F_{1}(g_{1})-F_{1}(g_{2})+F_{0}]\mu
(g_{1})\mu (g_{2})dxdg_{1}dg_{2}+...,  \notag
\end{align}%
where now the symbol $\langle ...\rangle $ in the l.h.s. of this formula
denotes averaging with respect to $\{\xi _{j}\}$ and $\{g_{j}\}$ and
\begin{equation}
\mu (g)=\mu p(g).  \label{mug}
\end{equation}

\subsection{Density expansion of the DOS}

Now we apply the expansion described above to the density of states of the
Schr\"{o}dinger equation. We use the self-averaging property of the DOS,
according to which \cite{LGP}
\begin{equation}
\rho (E)=\lim_{\Lambda \rightarrow \infty }\langle |\Lambda
|^{-1}\sum_{n\geq 1}\delta (E-E_{n})\rangle ,  \label{dosa}
\end{equation}
where $\{E_{n}\}_{n\geq 1}$ are the energy levels of the Hamiltonian $%
H_{\Lambda }$ defined by the Schr\"{o}dinger equation with the potential (%
\ref{Veff1}) in the cube $\Lambda $.

Comparing the l.h.s. of (\ref{exp2}) and the r.h.s. of (\ref{dosa}), we
conclude that in this case the role of $F_{l}$ in (\ref{exp2}) play
\begin{equation*}
\sum_{n\geq 1}\delta (E-E_{n}^{(l)}((x_{1},g_{1}),...,(x_{l},g_{l}))),
\end{equation*}%
where $\{E_{n}^{(l)}((x_{1},g_{1}),...,(x_{l},g_{l}))\}_{n\geq 1}$ is the
negative spectrum of the $l$-wells Hamiltonian
\begin{equation}
H^{(l)}=-\Delta +\sum_{j=1}^{l}g_{j}v\left( \sqrt{g_{j}}(x-x_{j})\right) .
\label{Hl}
\end{equation}%
Thus, applying (\ref{exp2}) to the DOS and taking into account that we are
interested in negative energies of large absolute value and that $%
H^{(0)}=-\Delta $ has no negative spectrum, we find that the term $\rho
^{(0)}(E)$ with $l=0$ (the zero-well contribution) is absent in the
expansion. Hence, the leading contribution in $\mu $ to the DOS is due to
the one-well term of the expansion:
\begin{equation}
\rho ^{(1)}(E)=\sum_{n\geq 1}\int \delta (E-E_{n}^{(1)})\mu (g)dg.
\label{rho1}
\end{equation}%
For the well of the form $gv\left( \sqrt{g}x\right) $ we have:
\begin{equation}
E_{n}^{(1)}=g\varepsilon _{n},  \label{Eg1}
\end{equation}%
where $\{\varepsilon _{n}\}_{n\geq 1}$ are the negative eigenvalues of the
dimensionless operator $-\Delta +v(x)$. Thus
\begin{equation*}
\rho ^{(1)}(E)=\sum_{n\geq 1}\mu \left( \frac{E}{\varepsilon _{n}}\right)
\frac{1}{|\varepsilon _{n}|}.
\end{equation*}%
According to the spirit of our approach the density $p(g)$ should decay
sufficiently fast as $g\rightarrow \infty $. Thus the leading contribution
to $\rho ^{(1)}(E)$ is due to the first term of the sum, i.e., we can use
the approximation
\begin{equation}
\rho ^{(1)}(E)\simeq \mu \left( \frac{E}{\varepsilon _{1}}\right) \frac{1}{%
|\varepsilon _{1}|}.  \label{rhome}
\end{equation}%
Normalizing the well $v$ by the condition
\begin{equation}
\varepsilon _{1}=-1,  \label{e0}
\end{equation}%
we can write
\begin{equation}
\rho ^{(1)}(E)\simeq \mu (-E).  \label{rhomu}
\end{equation}%
The last formula is a version of the well known \textquotedblleft
classical\textquotedblright\ asymptotic formula for the DOS valid for smooth
random potentials. By choosing as a randomizing parameter of the wells $v_{j}
$ in (\ref{Veff}) their ground state energies, we can show that an analogue
of (\ref{rhome}) allows us to obtain also "quantum" versions of asymptotic
formulas for the DOS valid for singular $v$'s (see \cite{LGP} for the
respective terminology and results).

It can also be shown that the two-well contribution to the DOS is
of the order $O(\mu ^{2})$. We postpone the corresponding argument
to Section 5.1. Thus the two-well contribution is negligible with
respect to the r.h.s.
of (\ref{rhomu}). We conclude that the unknown (and small) function $\mu (g)$%
, determining our effective potential and having the sense of the
probability density to find a well of amplitude lying between $g$ and $g+dg$
with center in an infinitesimal neighborhood of a given $x$, coincides in
our approximation with the DOS of the Schr\"{o}dinger operator. This
important conclusion makes our scheme self-consistent. It corresponds to the
basic ingredient of the Lifshitz approach, according to which the DOS is the
probability density of the localization wells, having the ground state
energy $E$ \cite{Li:65}. This interpretation of the DOS is widely used in
the theory of disordered systems \cite{Ef-Sh:84,LGP}. In our approach it is
a simple consequence of the ansatz (\ref{Veff1}) and of the expansion
formulas of the previous section.

Let $\Delta $ be an interval of values of random variables $g_{j}$, lying in
the strong localized spectrum with width much smaller than typical values of
the $g$'s under consideration. Then $\bar \mu =\int_{\Delta }\mu (g)dg$ will
be the density per unit volume of wells, whose amplitudes are in $\Delta $,
and $\bar \mu ^{-1/d}$ will be the typical distance between these wells. Our
approach is based on the assumption that typical distances between wells are
much larger than the typical radii of the localization wells (cf (\ref{cond}%
)). In the case of the effective potential (\ref{Veff1}), this assumption
can be written as
\begin{equation}
g^{-1/2}\ll \bar \mu ^{-1/d}.  \label{cond1}
\end{equation}

\section{A.C. Conductivity}

\label{s:cond}

\subsection{Generalities}

\label{ss:gener}

Recall that from the point of view of statistical physics, we are dealing
with an ideal gas of electrons in the external random field $V(x)$ (one-body
approximation). In this case, the linear response theory leads to the
following formula for the tensor of the zero-temperature a.c. conductivity
of a macroscopic system of spinless electrons in an external spatially
homogeneous electric field of the frequency $\nu $ at zero temperature:
\begin{equation*}
\sigma _{\alpha \beta }(\nu ,E_{F})=\lim_{\Lambda \rightarrow \infty }\pi
e^{2}|\Lambda |^{-1}\sum_{m\neq n}\delta (E_{F}+\nu -E_{m})\delta
(E_{F}-E_{n})V_{mn}^{(\alpha )}V_{nm}^{(\beta )},
\end{equation*}%
where $V_{mn}^{(\alpha )}$ are the matrix elements of the velocity operator $%
i\triangledown _{\alpha }$ between the states $\psi _{m}$ and $\psi _{n}$ of
the system, confined to the box $\Lambda $. In the case of a random
potential, homogeneous in mean and weakly correlated, the conductivity is
self-averaging \cite{LGP}. Thus we have in the thermodynamic limit, assuming
for simplicity that the system is rotational invariant in mean:
\begin{equation}
\sigma _{\alpha \beta }(\nu ,E_{F})=\frac{\pi e^{2}}{d}\delta _{\alpha \beta
}\sigma (\nu ,E_{F}),  \label{Ku01}
\end{equation}%
\begin{equation}
\sigma (\nu ,E_{F})=\lim_{\Lambda \rightarrow \infty }\left\langle |\Lambda
|^{-1}\sum_{m\neq n}\delta (E_{F}+\nu -E_{m})\delta
(E_{F}-E_{n})|V_{mn}|^{2}\right\rangle ,  \label{Ku1}
\end{equation}%
where
\begin{equation}
|V_{mn}|^{2}=\sum_{\alpha =1}^{d}|V_{mn}^{(\alpha )}|^{2}.  \label{Vmna2}
\end{equation}%
Since $V=i[H,X]$, where $X$ is the coordinate operator, we have $%
|V_{mn}^{(\alpha )}|=|(E_{m}-E_{n})X_{mn}^{(\alpha )}|$, and (\ref{Ku1}) can
be written as
\begin{equation}
\sigma (\nu ,E_{F})=\nu ^{2}\lim_{\Lambda \rightarrow \infty }\langle
|\Lambda |^{-1}\sum_{m\neq n}\delta (E_{F}+\nu -E_{m})\delta
(E_{F}-E_{n})|X_{mn}|^{2}\rangle .  \label{Ku2}
\end{equation}%
where
\begin{equation}
|X_{mn}|^{2}=\sum_{\alpha =1}^{d}|X_{mn}^{(\alpha )}|^{2}.  \label{Xmna2}
\end{equation}%
Note that we keep the frequency $\nu $ non-zero while making the
thermodynamic limit $\Lambda \rightarrow \infty $ in the above formulas.
%Otherwise the macroscopic d.c. conductivity would be zero.
This prescription is well known in kinetic theory and is reminiscent of
keeping non-zero magnetic field while making the thermodynamic limit for a
ferromagnetic system in order to obtain non-zero macroscopic spontaneous
magnetization. Another way to obtain non-zero d.c. conductivity is to set $%
\nu =0$ in (\ref{Ku1}) but to replace the $\delta $-functions by a
sharp function of width $\eta $ (usually by the Lorenzian). This
corresponds to an imaginary shift in energies instead of a
real-valued shift $\nu $ (see e.g. \cite{Ai-Gr}, where the an
imaginary shift is used). In this paper, we will use the formula
(\ref{Ku2}), assuming always that the frequency is non-zero, but
small compared to the Fermi energy, i.e., we will assume that inequality (%
\ref{cond}) holds.

\subsection{Computation}

\label{ss:comp}

Now we are going to apply the density expansion formula (\ref{exp2}) to the
a.c. conductivity. Comparing (\ref{Ku2}) and (\ref{exp2}), we choose the
functions $F_{l}$ in this case as
\begin{equation}
\nu ^{2}\sum_{m\neq n}\delta (E_{F}+\nu -E_{m}^{(l)})\delta
(E_{F}-E_{n}^{(l)})|X_{mn}^{(l)}|^{2},  \label{Flc}
\end{equation}%
where $\{E_{n}^{(l)}\}_{n\geq 1}$ are negative levels of the $l$-wells
Hamiltonian (\ref{Hl}), and
\begin{equation}
X_{mn}^{(l)}=\int x\psi _{m}^{(l)}(x)\psi _{n}^{(l)}dx,  \label{Xl}
\end{equation}%
.Here the $\{\psi _{n}^{(l)}\}_{n\geq 1}$ are the bound states of (\ref{Hl}%
). By the same reason as in the case of the DOS, the zero-well contribution $%
\sigma ^{(0)}$ to the conductivity expansion is absent. Let us show that the
one-well contribution $\sigma ^{(1)}$ is also absent. Combining (\ref{Flc})
for $l=1$ and (\ref{Eg1}), we obtain
\begin{equation*}
\sigma ^{(1)}(\nu ,E_{F})=\nu ^{2}\sum_{m\neq n}\int \delta (E_{F}+\nu
-g\varepsilon _{m})\delta (E_{F}-g\varepsilon _{n})|X_{mn}^{(1)}|^{2}\mu
(g)dg.
\end{equation*}%
where $\{E_{n}^{(1)}\}_{n\geq 1}$ are the bound state energies (\ref{Eg1})
of the one-well Hamiltonian $H^{(1)}(g)=-\Delta +gv(\sqrt{g}x)$, and $%
X_{mn}^{(1)}$ is the coordinate matrix element between the corresponding
states $\{\psi _{l}^{(1)}\}$. Non-zero contributions to this expression are
due to the pairs $(m,n)$ such that
\begin{equation}
g\varepsilon _{n}=E_{F},\ g\varepsilon _{m}=E_{F}+\nu \simeq E_{F},\
g(\varepsilon _{n}-\varepsilon _{m})=\nu .  \label{encons}
\end{equation}%
Denoting by $\varepsilon $ the typical value of the levels $\varepsilon _{n}$%
's of the potential well $v$ and by $\delta \varepsilon $ the typical value
of the spacings $|\varepsilon _{n+1}-\varepsilon _{n}|$, we see that the
above conditions are incompatible if $g\delta \varepsilon \gg \nu $, i.e.,
if $E_{F}\delta \varepsilon /\varepsilon \gg \nu $. Since $\varepsilon _{n}$%
's are dimensionless, the last condition is just another form of our basic
condition (\ref{cond}).

The two level contribution $\sigma ^{(2)}(\nu ,E_{F})$ to the a.c.
conductivity is (cf (\ref{exp2})):
\begin{align}
\sigma ^{(2)}(\nu ,E_{F})& =\frac{\nu ^{2}}{2}\sum_{m\neq n}\int \delta
(E_{F}+\nu -E_{m}^{(2)})\delta (E_{F}-E_{n}^{(2)})  \label{si2} \\
& \times |X_{mn}^{(2)}|^{2}\mu (g_{1})\mu (g_{2})dg_{1}dg_{2}dy,  \notag
\end{align}
where $\{E_{n}^{(2)}\}_{n\geq 1}$ are the bound state energies of the
two-well Hamiltonian
\begin{equation}
H^{(2)}((x_{1},g_{1}),(x_{2},g_{2}))=-\Delta +g_{1}v_{1}+g_{2}v_{2},
\label{H2}
\end{equation}
in which
\begin{equation*}
v_{k}(x)=v\left( \sqrt{g_{k}}(x-x_{k})\right) ,\;k=1,2,
\end{equation*}
$y=x_{1}-x_{2}$, and $X_{mn}^{(2)}$ are the corresponding coordinate matrix
elements.

In view of our basic condition (\ref{cond1}), we have typically $%
|x_{1}-x_{2}|\gg \mathrm{max}\ g_{1,2}^{-1/2}$. Hence, according to general
principles of quantum mechanics, each level of (\ref{H2}) should be
(exponentially) close to a certain level of one of infinite distant wells,
and each eigenfunction is (exponentially) close either to an eigenfunction
of one of the wells (non-resonant case) or to a linear combination of the
eigenfunctions of the both wells with coefficients of the same order of
magnitude (resonant case).

To make this description more quantitative, consider the one-well
Hamiltonians
\begin{equation*}
H_{k}^{(1)}=-\Delta +g_{k}v_{k},\ k=1,2,
\end{equation*}%
corresponding to (\ref{H2}). Normalize the potential well $v(x)$ by the same
condition (\ref{e0}). Then the lowest eigenvalues of $H_{k}^{(1)},\;k=1,2$
are $-g_{k}$, and the corresponding eigenfunctions are
\begin{equation}
\varphi _{k}(x)=g_{k}^{d/4}\varphi \left( \sqrt{g_{k}}(x-x_{k})\right) ,\
k=1,2,  \label{phi}
\end{equation}%
where $\varphi (x)$ is the ground state of the dimensionless operator $%
-\Delta +v(x)$. The function $\varphi (x)$ decays exponentially in $x$ with
rate 1. Hence
\begin{equation}
\varphi _{k}(x)\sim \exp \left( -\sqrt{g_{k}}|x-x_{k}|\right) ,\
|x-x_{k}|\gg g_{k}^{-1/2}.  \label{phias}
\end{equation}%
Since we will be interested mostly in the resonant case, we assume that $%
g_{1,2}\simeq |E_{F}|$, i.e., the radii of the $\varphi _{k},\ k=1,2$ in (%
\ref{phi}) are of the same order of magnitude
\begin{equation}
g_{1,2}^{-1/2}\simeq r_{l}=|E_{F}|^{-1/2},  \label{rl}
\end{equation}%
hence $r_{l}\ll |x_{1}-x_{2}|$.

In this situation we can find the lowest eigenvalues of $H^{(2)}$ in the
framework of the widely used approximation, in which $H^{(2)}$ is replaced
by its projection on the span of the functions $\varphi _{1}$ and $\varphi
_{2}$ \footnote{%
In the appendix, we compute exactly the negative spectrum of $H^{(2)}$ for $%
v(x)=-\delta (x)$ in the 1-dimensional case. The results for the
conductivity, obtained from this spectrum, coincide with those
found by using this approximation.}. The diagonal entries of this
$2\times 2$ matrix are
\begin{align*}
(\varphi _{k},H^{(2)}\varphi _{k})& =-g_{k}+g_{j\neq k}\int v_{j}(x)\varphi
_{k}^{2}(x)dx \\
& =-g_{k}+O\left( \exp \left( -2|x_{1}-x_{2}|/r_{l}\right) \right)
,\;|x_{1}-x_{2}|>>r_{l},
\end{align*}%
and its off-diagonal entry is
\begin{equation}
(\varphi _{1},H^{(2)}\varphi _{2})=-g_{1}(\varphi _{1},\varphi
_{2})+(\varphi _{1},v_{2}\varphi _{2}).  \notag
\end{equation}%
Since $v$ is of finite range, the first term here decays in $|x_{1}-x_{2}|$
not faster than the second term. Hence, being interested in distances $%
|x_{1}-x_{2}|$ that are much bigger than $g_{1,2}^{-1/2}$, we can neglect
the second term, i.e., we can use as the off-diagonal entry of the matrix
the quantity $-I(x_{1}-x_{2})$, where
\begin{equation}
I(x_{1}-x_{2})=g_{1}(\varphi _{1},\varphi _{2})\backsimeq g_{2}(\varphi
_{1},\varphi _{2})  \label{over}
\end{equation}%
is known as the overlap integral, and in view of (\ref{phi}) and (\ref{rl})
we have
\begin{equation}
I(x)\simeq I_{0}e^{-|x|/r_{l}},\ |x|\gg r_{l},  \label{Ias}
\end{equation}%
with
\begin{equation}
I_{0}\simeq |E_{F}|.  \label{Io}
\end{equation}%
We obtain that the two lowest eigenvalue of the two-well Hamiltonian $H^{(2)}
$ can be found as the eigenvalues of the matrix
\begin{equation}
\left(
\begin{array}{cc}
-g_{1} & -I(x_{1}-x_{2}) \\
-I(x_{1}-x_{2}) & -g_{2}.%
\end{array}%
\right) .  \label{mat2}
\end{equation}%
Assuming that $g_{1}>g_{2}>0$, we obtain that the eigenvalues of this matrix
are
\begin{equation}
E_{k}^{(2)}=-g-(-1)^{k-1}\sqrt{\delta ^{2}+I^{2}},\ k=1,2,  \label{levs}
\end{equation}%
where
\begin{equation}
g=\frac{g_{1}+g_{2}}{2},\ \delta =\frac{g_{1}-g_{2}}{2},  \label{deg}
\end{equation}%
and the corresponding eigenfunctions of the projection of $H^{(2)}$ are
\begin{equation}
\begin{array}{l}
\psi _{1}(x)=\varphi _{1}(x)\cos \theta +\varphi _{2}(x)\sin \theta  \\
\psi _{2}(x)=-\varphi _{1}(x)\sin \theta +\varphi _{2}(x)\cos \theta ,%
\end{array}
\label{psi}
\end{equation}%
where
\begin{equation}
\tan \theta =\frac{I}{\delta +\sqrt{\delta ^{2}+I^{2}}}.  \label{tan}
\end{equation}%
%
%
%The last formulas show that the condition of the resonant tunnelling is
%\begin{equation}
%\delta \ll |I(x_{1}-x_{2})|,  \label{res}
%\end{equation}%
%or, in view of (\ref{Ias}) - (\ref{Io}), that $|x_{1}-x_{2}|\gg r_{l}\log
%I_{0}/\delta \sim g^{-1/2}\log I_{0}/\delta \gg r_{l}$. In this case
%\begin{equation}
%\psi _{n}=\frac{\varphi _{1}+(-1)^{n-1}\varphi _{2}}{\sqrt{2}},\ n=1,2,
%\label{psir}
%\end{equation}%
%i.e., a quantum particle is equally shared by two wells up to the
%exponentially small corrections in $|x_{1}-x_{2}|/r_{l}$. Besides,
%it is
%easy to see that if $|x_{1}-x_{2}|\gg r_{l}\sim g^{-1/2}$, then the norm of $%
%\psi _{1,2}$ is asymptotically close to $1$, because
%\begin{equation}
%||\psi _{1,2}||^{2}=1\pm \sin 2\theta \int \varphi _{1}\varphi
%_{2}dx=1+O(e^{-\sqrt{g}|x_{1}-x_{2}|}).  \label{norm}
%\end{equation}%
%We shall see below that the values of $|x_{1}-x_{2}|$, contributing to the
%conductivity and to the pair correlators, correspond to the resonance
%tunnelling regime (\ref{res}), in which $\delta \leq \nu /2$. Hence, in view
%of (\ref{Ias}) - (\ref{Io}), the corrections to (\ref{psir}) and to the unit
%norm of $\psi _{1,2}$ are of the exponential order in $\log I_{0}/\nu \gg 1$%
%, i.e., of the order $\nu /|E_{F}|\ll 1$, and will be neglected in what
%follows.
%These are familiar forms of the states and levels of a two-well
%potential in the case, where the infinite distant wells have
%almost coinciding levels.
We are going to use these formulas in the r.h.s. of (\ref{si2}), keeping
there only terms with $m,n=1,2$, i.e., in fact, the term, corresponding to $%
m=1,\;n=2$. It is easy to see that the equalities $E_{F}=E_{1}^{(2)},%
\;E_{F}+\nu =E_{2}^{(2)}$ imply, in view of (\ref{levs})--(\ref{deg}), that $%
\nu =2\sqrt{I^{2}(y)+\delta ^{2}},\;y=x_{2}-x_{1}$. Hence, by (\ref{Ias})--(%
\ref{Io}) and by the condition $\nu \ll |E_{F}|$, the values of $y$,
contributing to (\ref{si2}), are bounded below by
\begin{equation}
r(\nu )=r_{l}\log \frac{2I_{0}}{\nu },  \label{rnu}
\end{equation}%
and the values of $|\delta |$ do not exceed $\nu /2$. Under these conditions
the coordinate matrix element $X_{12}^{(2)}$ in (\ref{si2}):
\begin{align}
X_{12}^{(2)}& =(x_{1}-x_{2})\frac{I}{2\sqrt{\delta ^{2}+I^{2}}}%
+(g_{1}^{-1/2}-g_{2}^{-1/2})\frac{I}{2\sqrt{\delta ^{2}+I^{2}}}\int x\varphi
^{2}(x)dx  \label{X12} \\
& +\frac{\delta }{\sqrt{\delta ^{2}+I^{2}}}\int x\varphi _{1}(x)\varphi
_{2}(x)dx
\end{align}%
between states (\ref{psi}) can be replaced by
\begin{equation}
X_{12}^{(2)}\simeq (x_{1}-x_{2})\frac{I}{2\sqrt{\delta ^{2}+I^{2}}}.
\label{x10}
\end{equation}%
Indeed, the second term in (\ref{X12}) can be omitted because its ratio to
the first term is of the order $\nu (|E_{F}|\log 2|E_{F}|/\nu )^{-1}\ll 1$.
Besides, the term is zero if $\varphi $ is even. The relative order of the
third term is the same as the second one.

In view of the above we obtain that the two-well contribution (\ref{si2}) to
the a.c. conductivity is
\begin{equation}
\sigma ^{(2)}(\nu ,E_{F})=\nu \int \frac{|y|^{2}I^{2}(y)}{\delta
^{2}+I^{2}(y)}\delta (E_{F}+\nu -E_{2}^{(2)})\delta (E_{F}-E_{1}^{(2)})\mu
(g_{1})\mu (g_{2})dg_{1}dg_{2}dy.  \label{simu}
\end{equation}%
We integrate first the product of two $\delta $-functions with respect to $%
g_{1}$ and $g_{2}$, taking into account that $|g_{1}-g_{2}|\lesssim \nu \ll
|E_{F}|\sim g_{1,2}$. This allows us to replace $\mu (g_{1})$ and $\mu
(g_{2})$ by $\mu (-E_{F})$, to set
\begin{equation}
\delta =\frac{1}{2}\sqrt{\nu ^{2}-4I^{2}(y)},  \label{dnI}
\end{equation}%
and to obtain in view of (\ref{rhomu})
\begin{equation}
\sigma ^{(2)}(\nu ,E_{F})=\nu \rho ^{2}(E_{F})\int_{2|I(y)|\geq \nu }\frac{%
|y|^{2}I^{2}(y)}{\sqrt{\nu ^{2}-4I^{2}(y)}}dy.  \label{siI}
\end{equation}%
Note that the restriction $2|I(y)|\geq \nu $ of the domain of integration in
(\ref{siI}) is because of the presence of the two $\delta $-functions in (%
\ref{simu}), i.e., in fact, because of energy conservation.

In view of the inequalities $0<\nu <<E_{F}$ and formulas (\ref{Ias})--(\ref%
{Io}), we can replace the condition $2|I(y)|\geq \nu $ by the condition $%
|y|\geq r(\nu )$, where $r(\nu )$ is defined in (\ref{rnu}).
%\begin{equation}
%I_{0}e^{-r/r_{l}}=\nu /2,  \label{res1}
%\end{equation}
%i.e.,
%\begin{equation}
%r(\nu )=r_{l}\log \frac{2I_{0}}{\nu }\gg r_{l}.  \label{rnu}
%\end{equation}
%Formulas (\ref{levs}) - (\ref{psir}) allow us to interpret the last
%relations as conditions for the two ''bare'' states $\varphi _{1,2}$ to be
%in resonance.

The integrand in (\ref{siI}) is divergent at the lower limit $|y|=r(\nu )$
and decays exponentially fast at infinity with the rate $2/r_{l}$ in view of
(\ref{Ias}). Thus the main contribution to the integral is due to a $r_{l}$%
-neighborhood of the lower integration limit. This leads to the asymptotic
expression
\begin{equation}
\sigma ^{(2)}(\nu ,E_{F})=\frac{\nu ^{2}\rho ^{2}(E_{F})S_{d}}{4}%
r_{l}^{d+2}\left( \log \frac{2I_{0}}{\nu }\right) ^{d+1}.  \label{sir1}
\end{equation}%
where $S_{d}$ is the area of the $d$-dimensional sphere. Taking into account
relations (\ref{Io}), and (\ref{rl}), we obtain finally that
\begin{equation}
\sigma ^{(2)}(\nu ,E_{F})=\frac{\nu ^{2}\rho ^{2}(E_{F})S_{d}}{4}%
|E_{F}|^{-(d+2)/2}\left( \log \frac{2|E_{F}|}{\nu }\right) ^{d+1}.
\label{sif}
\end{equation}%
In particular, we have for $d=1$:
\begin{equation}
\sigma (\nu ,E_{F})=\frac{\nu ^{2}\rho ^{2}(E_{F})}{2}|E_{F}|^{-3/2}\left(
\log \frac{2|E_{F}|}{\nu }\right) ^{2}.  \label{si1}
\end{equation}%
These are our versions of the Mott formula (\ref{Mott}). They will be
discussed in more details in Section 5.

\section{Correlation Functions}

\subsection{Generalities}

In this section we study the following two-point correlation functions:
\begin{equation}  \label{C1}
C_{1}(x-y;\nu ,E)=\langle \rho _{E}(x,y)\rho _{E+\nu }(y,x)\rangle ,
\end{equation}
and
\begin{equation}  \label{C2}
C_{2}(x-y;\nu ,E)=\langle \rho _{E}(x,x)\rho _{E+\nu }(y,y)\rangle ,
\end{equation}
In writing the above expressions, we took into account the translation
invariance in coordinates of the correlation functions, following from the
translation invariance in mean, a fundamental property of disordered systems.

The function $C_{1}$ of (\ref{C1}) is closely related to the a.c.
conductivity. Indeed, recall the spectral theorem, according to which
\begin{equation}
\rho _{E}(x,y)=\int \delta (E-E^{\prime })\psi _{E^{\prime }}(x)\psi
_{E^{\prime }}(y)dE^{\prime },  \label{spth}
\end{equation}
where the symbol $\int ...dE$ denotes both the integration over the
continuous spectrum and the summation over the point spectrum.

Formulas (\ref{spth}), (\ref{Ku2}), and (\ref{C1}) imply that
\begin{equation}
\sigma (\nu ,E)=-\frac{\nu ^{2}}{2}\int |x|^{2}C_{1}(x,E,\nu )dx.
\label{siC1}
\end{equation}
The function $C_{2}$ of (\ref{C2}) is the local DOS--DOS correlator and is a
characteristic of localization, providing information on correlations of
eigenstates whose energy difference is $\nu $ and that are localized in
spatial domains of distance $x-y$.

Comparing (\ref{Kl}) for $l=2$ and (\ref{C1}) and (\ref{C2}), we obtain the
equalities
\begin{equation}
\begin{array}{l}
C_{1}(x;\nu ,E)=K_{2}(0,x;x,0;E,E+\nu ), \\
C_{2}(x;\nu ,E)=K_{2}(0,0;x,x;E,E+\nu ).%
\end{array}
\label{cK}
\end{equation}
We list below certain properties of $C_{1}$ and $C_{2}$.

\medskip \noindent (i)
\begin{equation}
|C_{1}(x;\nu ,E)|\leq C_{2}(x;\nu ,E).  \label{ci}
\end{equation}
The inequality follows from the inequality $|\rho _{E}(x,y)|^{2}\leq \rho
_{E}(x,x)\rho _{E}(y,y)$ that is a simple consequence of the Schwarz
inequality $\left\langle ab\right\rangle ^{2}\leq \left\langle
a^{2}\right\rangle \left\langle b^{2}\right\rangle $ and of the spectral
theorem (\ref{spth}).

\medskip \noindent (ii)
\begin{equation}
\int C_{1}(x;\nu ,E)dx=\delta (\nu )\rho (E).  \label{Cii}
\end{equation}
This relation follows from (\ref{C1}) and (\ref{dos}) and can be interpreted
as a weak form of the decay of the correlator $C_{1}$ at infinity.

\medskip \noindent (iii)
\begin{equation}
\lim_{\Lambda \rightarrow \infty }|\Lambda |^{-1}\int_{\Lambda }C_{2}(x;\nu
,E)dx=\rho (E)\rho (E+\nu ).  \label{Ciii}
\end{equation}%
To prove this formula, we use the ergodic theorem for $\rho _{E}(x,x)$,
implying the validity of the relation
\begin{equation*}
\lim_{\Lambda \rightarrow \infty }|\Lambda |^{-1}\int_{\Lambda }\rho
_{E}(x,x)dx=\langle \rho _{E}(0,0)\rangle \equiv \rho (E)
\end{equation*}%
on almost all realizations of the random potential. Applicability of the
ergodic theorem follows from the translation invariance in mean and the
decay of the spatial correlation in disordered systems (see e.g. \cite{LGP}).

Formula (\ref{Ciii}) expresses the decay of correlations between two density
operators in (\ref{C2}) as $|x_{1}-x_{2}|\rightarrow \infty $. Indeed, its
r.h.s. is the product of the averages of these two operators (see (\ref{dos}%
)), and its l.h.s. is a weak form of the relation $\lim_{x\rightarrow \infty
}C_{2}(x,E,E+\nu )=\rho (E)\rho (E+\nu )$.

\medskip \noindent (iv) Assume that for a certain $E$
\begin{equation}
C_{\alpha }(x;\nu ,E)=\delta (\nu )p_{\alpha }(x;E),\ \alpha =1,2.
\label{Cpa}
\end{equation}
Then

\begin{itemize}
\item[(a)]
\begin{equation*}
p_{1}(x;E)=p_{2}(x;E)= p(x;E)\geq 0;
\end{equation*}

\item[(b)]
\begin{equation}
p(x;E)=\left\langle \sum_{\text{loc}}\delta (E-E_{j})\psi _{j}^{2}(0)\psi
_{j}^{2}(x)\right\rangle ,  \label{And}
\end{equation}
where the symbol $\sum_{\text{loc}}$ denotes the summation over the
localized states only;

\item[(c)] if one defines the density of localized states as
\begin{equation*}
\rho _{loc}(E) = \int p(x;E)dx=\left\langle \sum_{\mathrm{loc}}\delta
(E-E_{j})\psi _{j}^{2}(0)\right\rangle ,
\end{equation*}%
then
\begin{equation}
\rho _{\mathrm{loc}}(E)\leq \rho (E),  \label{rlr}
\end{equation}%
and the inequality $\rho _{\mathrm{loc}}(E)>0$ is equivalent to the
existence of localized states in a neighborhood of $E$, and the equality $%
\rho _{\mathrm{loc}}(E)=\rho (E)$ is equivalent to complete localization in
a neighborhood of $E$.
\end{itemize}

\smallskip \noindent The above properties follow from the spectral theorem (%
\ref{spth}). The functions $p_{\alpha }(x;E)$ of (\ref{Cpa}) are the
\textquotedblright diagonal parts\textquotedblright\ of the r.h.s. of
equalities (\ref{cK}), viewed as functions of two variables $E_{1}=E$ and $%
E_{2}=E+\nu $.

The property (iv) will not be used below. We presented this property to
demonstrate usefulness of the correlators $C_{1}$ and $C_{2}$ in the theory
of disordered system. In particular, in the classic paper by P. Anderson
\cite{An:58} the positivity of $\int p(0;E)dE$ was used as an indicator for
localization. The quantum mechanical meaning of $\int p(0;E)dE$ is the
probability for a particle to be in an infinitesimal neighborhood of the
origin at time $t=\infty $, provided that at $t=0$ it was at the origin (the
return probability density)\cite{LGP}.

\subsection{Computations.}

To apply the density expansion formula (\ref{exp2}) to the correlation
functions (\ref{C1}) and (\ref{C2}), we write them in the form of extensive
quantities per unit volume:
\begin{equation}
C_{\alpha }(x)=|\Lambda |^{-1}\Phi _{\alpha }(x),\ \alpha =1,2,  \label{Cal}
\end{equation}%
where
\begin{equation}
\Phi _{1}(x)=\int_{\Lambda }C_{1}((x+a)-a)da=\int_{\Lambda }\langle \rho
_{E}(a,x+a)\rho _{E+\nu }(x+a,a)\rangle da,  \label{phi1}
\end{equation}%
and
\begin{equation}
\Phi _{2}(x)=\int_{\Lambda }C_{2}((x+a)-a)da=\int_{\Lambda }\langle \rho
_{E}(a,a)\rho _{E+\nu }(x+a,x+a)\rangle da,  \label{phi2}
\end{equation}%
Now it is clear that the role of the functions $F_{l}$ in (\ref{exp2}) for $%
C_{\alpha }$will play $\Phi _{\alpha }$, written for the $l$-well
Hamiltonian (\ref{Hl}).

By using these formulas, the zero-well and the one-well contributions to $%
C_{\alpha } $, $\alpha =1,2$ are absent by the same argument as for the
conductivity. The two-well contribution $C_{1}^{(2)}$ to $C_{1}$ is (cf (\ref%
{si2})):
\begin{align}
C_{1}^{(2)}(x;\nu ,E)& =\frac{1}{2}\sum_{m, n}\int \delta (E+\nu
-E_{m})\delta (E-E_{n})  \notag \\
& \times \psi _{m}(a)\psi _{m}(a+x)\psi _{n}(a)\psi _{n}(a+x)\mu (g_{1})\mu
(g_{2})da dy dg_{1}dg_{2},  \label{C12}
\end{align}
where $y$ is the separation between two wells, implicit in $\psi _{m,n}$ and
in $E_{m,n}$.

Since $\nu >0$, the diagonal part $\sum_{m=n}$ of the double sum is zero.
Moreover, as in the case of the conductivity, we restrict ourselves to the
two lowest levels of the spectrum of $H^{(2)}$ of (\ref{H2}), found in the
previous section in the framework of the projection method. This leaves the
term $m=1,n=2$ in the double sum of (\ref{C12}). By using (\ref{levs}), we
can integrate with respect to $g_{1}$ and $g_{2}$ the product of two $\delta
$-functions, fixing $g_{1}$ and $g_{2}$ by the relations $E+g+\sqrt{\delta
^{2}+I^{2}}=0,\;E+g+\nu -\sqrt{\delta ^{2}+I^{2}}=0$ . In view of the
condition $0<\nu \ll E$ we obtain, replacing $\mu (-g_{1})$ and $\mu (-g_{1})
$ by $\rho (E)$ in view of (\ref{rhomu}) (cf (\ref{siI})):
\begin{align}
C_{1}^{(2)}(x;\nu ,E)& =\rho ^{2}(E)\nu \int da\psi _{1}(a)\psi _{2}(a)
\label{C211} \\
& \times \int_{2|I(y)|\geq \nu }\psi _{1}(a+x)\psi _{2}(a+x)\frac{\nu }{%
\sqrt{\nu ^{2}-4I^{2}(y)}}dy.  \notag
\end{align}%
According to the previous section, if $\nu \ll E$, the restriction $%
2|I(y)|\geq \nu $ is equivalent to $|y|\geq r(\nu )$, where the resonant
radius $r(\nu )\gg r_{l}$ is defined in (\ref{rnu}). This and the form (\ref%
{psi}) of the functions $\psi _{1,2}$ imply the that if $|g_{1}-g_{2}|%
\lesssim \nu \ll g\sim |E|$, then
\begin{equation*}
\psi _{1}(a)\psi _{2}(a)=\cos \theta \sin \theta \left[ \varphi
^{2}(a)-\varphi ^{2}(a+y)\right] +O\left( e^{-2r(\nu )/r_{l}}\right) ,
\end{equation*}%
where $y=x_{1}-x_{2}$. The formula and the analogous formula with $a$,
replaced by $a+x$, lead to the following asymptotic expression for the
two-well contribution $C_{1}^{(2)}$ to the correlator $C_{1}$:
\begin{align}
C_{1}^{(2)}(x;\nu ,E)& =\frac{2\rho ^{2}(E)}{\nu }\int \varphi ^{2}(a)da
\label{C1I} \\
& \times \int_{|y|\geq r(\nu )}\frac{I^{2}(y)}{\sqrt{\nu ^{2}-4I^{2}(y)}}%
\left[ \varphi ^{2}(a+x)-\varphi ^{2}(a+x-y)\right] dy.  \notag
\end{align}%
Similar arguments show that the two-well contribution
\begin{align}
C_{2}^{(2)}(x;\nu ,E)& =\frac{1}{2}\sum_{m\neq n}\int \delta (E+\nu
-E_{m})\delta (E-E_{n})  \label{C220} \\
& \times \psi _{m}^{2}(a)\psi _{m}^{2}(a+x)\mu (g_{1})\mu
(g_{2})dadg_{1}dg_{2}dy,  \notag
\end{align}%
to the correlator $C_{2}$ is with the same accuracy:
\begin{align}
C_{2}^{(2)}(x;\nu ,E)& =C_{1}^{(2)}(x;\nu ,E)  \label{C2I} \\
& +\rho ^{2}(E)\int \varphi ^{2}(a)da\int_{|y|\geq r(\nu )}\frac{\nu }{\sqrt{%
\nu ^{2}-4I^{2}(y)}}\varphi ^{2}(a+x-y)dy.  \notag
\end{align}%
We formulate now several properties of $C_{1}^{(2)}$ and $C_{2}^{(2)}$,
following from (\ref{C1I})--(\ref{C2I}).

According to (\ref{C1I})
\begin{equation*}
\int C_{1}^{(2)}(x;\nu ,E)dx=0
\end{equation*}%
This relation is in agreement with the exact sum rule (\ref{Cii}), because
formula (\ref{C1I}) was obtained under the assumption that $\nu >0$.

Likewise, we have the limiting relation
\begin{equation}  \label{C22rho}
C_{2}^{(2)}\rightarrow \rho ^{2}(E),\;x\rightarrow \infty ,
\end{equation}
which is in agreement with the exact sum rule (\ref{Ciii}).

It is also easy to see that $C_{1}^{(2)}(x;\nu ,E)$:

\begin{enumerate}
\item[(i)] has a positive peak of order
\begin{equation}
\rho ^{2}(E)\left( \log \frac{2I_{0}}{\nu }\right) ^{d-1}  \label{peak}
\end{equation}
at the origin;

\item[(ii)] decays exponentially fast with the rate $2/r_{l}$ for $|x|\gg
r_{l}$, and is exponentially small in the spatial domain $r_{l} \ll |x| \ll
r(\nu )=r_{l}\log 2I_{0}/\nu \gg r_{l}$;

\item[(iii)] has a negative peak of the same order of magnitude (\ref{peak})
at a $r_{l}$-neighborhood of $|x|=r(\nu )$;

\item[(iv)] decays exponentially for $|x|\gg r(\nu )$ with the rate $2/r_{l}
$, thereby complementing (\ref{Ciii}).
\end{enumerate}

This behavior of $C_{1}^{(2)}$ allows us to obtain the Mott formula (\ref%
{sir1}) from the relations (\ref{siC1}) and (\ref{C1I}).

The correlator $C_{2}^{(2)}$ has the same behavior as $C_{1}^{(2)}$ in $x$
till $|x|\lesssim r(\nu )$, in particular it is exponentially small in $x$
if $r_{l}\ll |x|\ll r(\nu )$. Then $C_{1}^{(2)}$ becomes asymptotically
equal $\rho ^{2}(E)$ in the domain $|x-r(\nu )|\lesssim r_{l}$ and it is
equal to $\rho ^{2}(E)$ for all $x$, $|x|\gg r(\nu )$ (see (\ref{C22rho})).
In view of spectral theorem one can expect $\rho _{E}(x,x)$ to be
proportional to $\psi _{E}^{2}(x)$ in the strong localization regime (cf (%
\ref{rhoM})). Then the factorization property (\ref{C22rho}) can be
interpreted as the statistical independence of the localized states of
energies close to each others with separation much bigger than $r(\nu )$. On
the other hand, the exponential smallness of $C_{1,2}$ for $r_{l}\gg |x|\gg
r(\nu )$ can be interpreted as a kind of strong correlation between states
close in energy, that are not sufficiently well separated in space. These
correlations can be viewed as a manifestation of a certain "repulsion" of
nearby levels in the sense that the probability that nearby levels are close
tends to zero as the level spacing tends to zero (see \cite{Go-El,Al-Sh})
for discussions of this property). Figures 1 and 2 show examples of graphs
of $C_{1,2}$,
\begin{figure}[!h]
\begin{center}
\label{correl-1d}
\includegraphics[width=16cm,height=5cm]{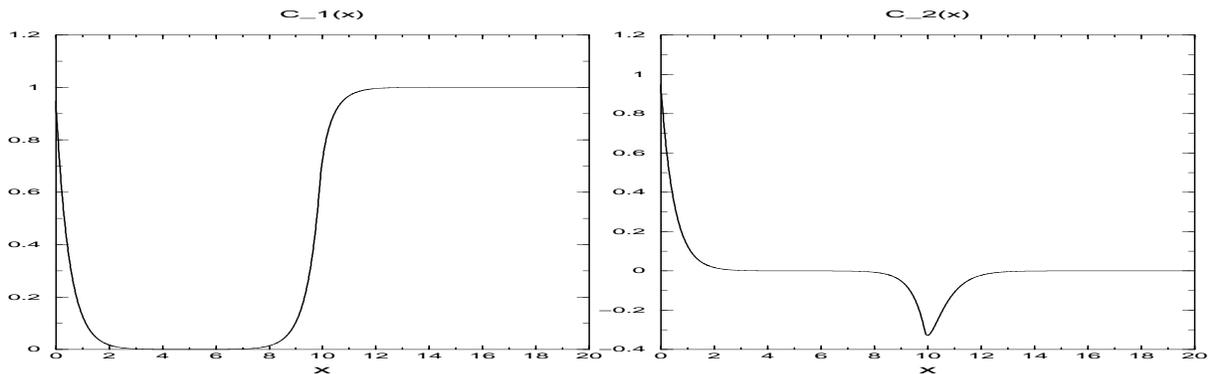}
\end{center}
\caption{{\protect\small {One dimensional correlation function $C_1(x)$ and $%
C_2(x)$ with $\protect\nu=10^{-4}$, $r_l=1$ and
$\protect\rho(E)=1$.}}}
\end{figure}
\ \newline
\newline

\begin{figure}[h]
\begin{center}
\label{correl-2d}
\includegraphics[width=16cm,height=5cm]{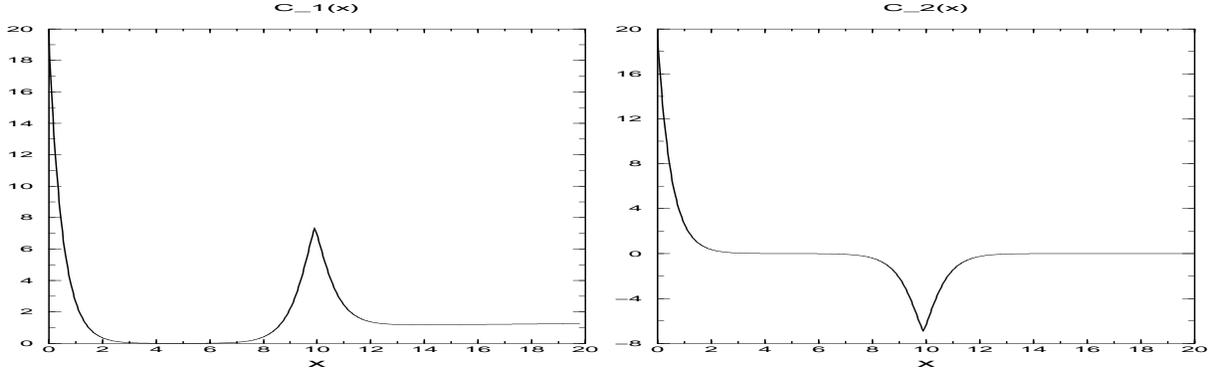}
\end{center}
\par
%\caption{\small{One dimensional correlation function $C_1(x)$ and $C_2(x)$
%with arbitrary constant $\nu=10^{-4}$, $r_l=1$ and $\rho(E)=1$.}}
\caption{{\protect\small {Two dimensional correlation function
$C_{1}(x)$ and $C_{2}(x)$ with $\protect\nu =10^{-4}$, $r_{l}=1$
and $\protect\rho (E)=1$.}}}
\end{figure}
Results similar to those outlined above were obtained as asymptotically
exact ones in \cite{Ha-Jo:91} in the one dimensional case of the strong
localization regime and in \cite{Go,Go-Co} in the one dimensional case of
the weak localization regime (see the following section for more details).
We see, however, that in dimension greater than 1 the characteristic value (%
\ref{peak}) of the peak of the correlation function (\ref{Ku1}) diverges as
the difference of the energies tends to zero. Thus, unlike the conductivity
that has the $\log $-factor in all dimensions, the correlation functions $%
C_{1,2}$ are logarithmically big in the energy difference for $|x|\sim r(\nu
)$ only in dimension bigger than 1. Similar results were obtained in \cite%
{We-Co} in the frameworks of the instanton approach (see Subsection 5.4).

The \textquotedblright two-hump\textquotedblright\ states (\ref{psi}) appear
in our approach just as a computational tool, allowing us to find leading
contributions to the low frequency conductivity and to the correlators $%
C_{1,2}$, by using the density expansion of Section 2.1, just as the
\textquotedblright one-hump\textquotedblright\ states (\ref{phi}) are
necessary to find the low energy asymptotic of the density of states in our
approach (see also Section 2.3), in the optimal fluctuation method \cite%
{Li:65,LGP,Ef-Sh:84}, and its version, known as the instanton approach (see
Section 5.4, \cite{We-Co} and references therein). On the other hand, the
development of localization theory of the last decades suggests that the
\textquotedblright one-hump\textquotedblright\ states carry certain
information on the structure of genuine localized states in disordered
systems. This suggests the belief, that the \textquotedblright
two-hump\textquotedblright\ states also reflect certain properties of
genuine localized states. If this is true, we can interpret the above
results on the spatial behavior of the correlators $C_{1,2}$ in the
following way. The existence of the length scale $r(\nu )$ of (\ref{rnu}),
that determines drastic changes of the spatial behavior of the correlators $%
C_{1,2}$, is due to the \textquotedblright
interaction\textquotedblright\ between close energy levels, and
the interaction mechanism is the resonant tunnelling between the
"bare" one-hump states, i.e., between different centers of genuine
states. The parameter $I_{0}$ of (\ref{Ias}) - (\ref{Io}) is the
characteristic interaction energy, determining the level splitting
(spacing), and $r(\nu )$ is the tunnelling distance, determined by
the two energy scales
($E=(E_{l}+E_{2})/2,\;v=|E_{2}-E_{1}|,\;E>>\nu $). This
inter-level interaction is a mechanism of a certain level
repulsion, that prevents the spatial domains where the states are
essentially non zero to be close and, as a result, leads to the
exponentially small values of the two-point correlators for
$r_{l}<<|x|<<r(\nu )$.

\section{Discussion}

\noindent \textit{5.1 Corrections}

\smallskip \noindent We comment now on the corrections (next terms of the
density expansions) to our formulas of Sections 2--4. We are not able to
prove the convergence of these expansions. We simply argue that they should
be asymptotic, i.e., that their terms should be small in successive powers
of $\rho (E)$. We will begin from the density of states itself.

It is easy to see that the next term in the expansion of the DOS has the
form
\begin{eqnarray*}
\rho ^{(2)}(E)& =&\int \left \{ \left[ \delta (E-E_{1}^{(2)})- \delta
(E-\varepsilon _{1}^{(2)})\right] \right. \\
& +&\left.\left[ \delta (E-E_{2}^{(2)})-\delta (E-\varepsilon _{2}^{(2)})%
\right] \right\} \mu (g_{1})\mu (g_{2})dydg_{1}dg_{2}.
\end{eqnarray*}
where $E_{1,2}^{(2)}$ are given by (\ref{levs}), and $\varepsilon
_{1}^{(2)}=-\max (g_{1},g_{2})$, $\varepsilon _{2}^{(2)}=-\min (g_{1},g_{2})$%
. Recall that we assume that $\mu (g)$ is smooth enough and decays
sufficiently fast for large $g$. Thus $\rho ^{(2)}$ will be of the order $%
O(\mu ^{2})$ if the integral in the relative distance $y$ between the wells
will be convergent. This fact follows from the inequality $%
|E_{k}-\varepsilon _{k}^{(2)}|\leq (\sqrt{\delta ^{2}+I^{2}(y)}-|\delta
|)\leq |I(y)|$, the exponential decay of $I(y)$ (see (\ref{Ias})), and the
smoothness of $\mu (g)$, allowing us to transfer derivatives of
delta-functions to the $\mu $'s.

In the general case of the correction of the order $l$ the appropriate
integrals in relative distances between wells will be convergent because of
the subtractions of the functions $F_{k}$ of lower orders $k<l$ from that of
the order $l$ in the $l$th term of the density expansion (\ref{exp2}), the
sufficiently fast splitting (additive clustering) of negative eigenvalues $%
E^{(l)}$ of the $l$-wells problem into the sums of negative eigenvalues $%
E^{(k)}$ of the $k$-wells problems $k<l$ and again because of the smoothness
of $\mu (g)$.

The situation is less simple in the case of the conductivity as we
have seen already for $l=2$. This is because of the presence of
families of tunnelling configurations for any number of wells (for
example, for $l=3$ there are two families: the equilateral
triangles and the three equidistant points on a straight line).
These configurations are responsible for the absence of decay (and
even for the polynomial growth) in distances between wells of
matrix elements $x_{ij}^{(l)}$ on the corresponding resonant
sub-manifolds and for the appearance of extra powers of $\log \nu
_{0}/\nu $ (where $\nu _{0}$ can be different from that of formula
(\ref{sif})). However, since the dimension of these resonant
manifolds grows slower than $l$, these powers of $\log \nu
_{0}/\nu $ will be always multiplied by powers of $\nu $, given by
the dimensions of the manifolds transversal to the resonant ones.
This is why the higher terms in the expansion of the low frequency
conductivity should be small compared to the terms in Mott's
formula (\ref{Mott}).

In other words, it seems reasonable to believe that these higher resonant
configurations will produce new peaks and new length scales in the higher
terms of the density expansion of the correlators, but that the amplitudes
of the peaks will be small relative to the amplitude (\ref{peak}) of the
peak due to the resonant pairs. One can also speculate that for bigger
densities of states (i.e., for energies closer to the mobility edge) higher
resonant configurations will play a more significant role, leading
eventually to the loss of the exponential decay of the correlators and to
the delocalization transition according to the scenario, outlined in \cite%
{Li:65,Th}

\medskip \noindent \textit{5.2. Asymptotically exact one-dimensional results}

\smallskip \noindent The asymptotic behavior of the low frequency
conductivity in the strong localization regime of the one-dimensional
Gaussian white noise potential, defined by the relations
\begin{equation}
\left\langle V(x)\right\rangle =0,\;\left\langle V(x)V(y)\right\rangle
=2D\delta (x-y),  \label{wn}
\end{equation}%
was studied in \cite{Ha-Jo:91}. The potential is often used in the theory of
one-dimensional disordered systems (see \cite{LGP} for results and
references). In particular, the density of states $\rho (E)$ and the
Lyapunov exponent $\gamma (E)$ of the Schr\"{o}dinger equation with this
potential can be found in quadratures. The strong localization regime
corresponds to negative energies of large absolute value
\begin{equation}
D^{2/3}\ll |E|.  \label{wnslr}
\end{equation}%
In this case we have the following asymptotic formulas \cite{LGP}
\begin{equation}
\rho (E)=\frac{2|E|}{\pi D}e^{-4|E|^{4/3}/3D},\;\gamma (E)=|E|^{1/2}.
\label{rhoga}
\end{equation}%
Moreover, the rate of the exponential decay of the eigenfunctions $\psi _{E}$
is $\gamma (E)$, because we have with probability 1 \cite{LGP,Pa-Fi:92}:
\begin{equation}
\lim_{|x|\rightarrow \infty }|x|^{-1}\log \left( \psi _{E}^{2}(x)+\psi
_{E}^{\prime 2}(x)\right) ^{1/2}=-\gamma (E)  \label{Lyap}
\end{equation}%
Hence, the exact asymptotic form of (\ref{rhoga}) for the localization
radius
\begin{equation}
r_{l}(E)=1/\gamma (E).  \label{rlga}
\end{equation}%
coincides with our approximate formula (\ref{rl}).

In the paper \cite{Ha-Jo:91} the low frequency conductivity was found using
the Grassmann functional integral representation of the Green's function,
which leads to an integral representation for the correlator $C_{1}$ of (\ref%
{C1}) (recall that the conductivity is related to the correlator via formula
(\ref{siC1})). The condition (\ref{wnslr}) allowed the authors to apply the
saddle point method to this integral representation. We will summarize the
results of \cite{Ha-Jo:91} in a form close to that of Sections 3 and 4.

The \textquotedblright two-hump\textquotedblright\ states, similar to (\ref%
{psi}) appear in \cite{Ha-Jo:91} as the saddle points of the effective
action for $C_{1}$. The states have in general a rather complicated
(two-instanton) form, but in the low frequency limit $0<\nu \ll |E|$ they
can be written in the form (\ref{psi}), in which the role of the
\textquotedblright bare\textquotedblright\ states play
\begin{equation}
\varphi _{1,2}(x)=\frac{1}{\sqrt{2r_{l}}\cosh \left( x\pm y/2\right) /r_{l}},
\label{fiHJ}
\end{equation}%
where $y\geq y_{0}(\nu )$ and
\begin{equation}
y_{0}(\nu )=r_{l}\log 8|E|/\nu   \label{rnus}
\end{equation}%
(cf (\ref{phi}), (\ref{phias}), and (\ref{rnu})). As for the angle $\theta $
of (\ref{psi}), it is defined by the relation $e^{-y/r_{l}}=e^{-y_{0}/r_{l}}%
\sin 2\theta $, that can be written as
\begin{equation}
\tan \theta =\frac{e^{-y/r_{l}}}{e^{-y_{0}/r_{l}}+\sqrt{%
e^{-2y_{0}/r_{l}}-e^{-2y/r_{l}}}}.  \label{tanHJ}
\end{equation}%
Introduce $\widetilde{I}(y)=\widetilde{I}_{0}e^{-y/r_{l}}$, where $%
\widetilde{I}_{0}=4|E|$. Then formula (\ref{rnus}) can be written as $%
\widetilde{I}(y_{0})=\nu /2$. These formulas have to be compared with (\ref%
{Ias}), and (\ref{Io}). Furthermore, setting
\begin{equation}
\widetilde{\delta }=\sqrt{\nu ^{2}/4-\widetilde{I}^{2}(y)}=\sqrt{\widetilde{I%
}^{2}(y_{0})-\widetilde{I}^{2}(y)},  \label{dera}
\end{equation}%
(cf (\ref{dnI})), we can write (\ref{tanHJ}) in a form, analogous to that of
(\ref{tan}).

According to \cite{Ha-Jo:91}, the correlator $C_{1}$ has the following
asymptotically exact form for $0 < \nu \ll |E|$:
\begin{equation*}
C_{1}(x;,E)=2\rho ^{2}(E)\int da\int_{y\geq y_{0}}\psi _{1}(a)\psi
_{1}(a+x)\psi _{2}(a)\psi _{2}(a+x)\frac{e^{-y/r_{l}}}{\sqrt{%
e^{-2y_{0}/r_{l}}-e^{-2y/r_{l}}}}dy,
\end{equation*}%
which can be written as (\ref{C211}) because, in view of the above
notations, we can write the expression $e^{-y/r_{l}}\left(
e^{-2y_{0}/r_{l}}-e^{-2y/r_{l}}\right) ^{-1/2}$ in the last formula as $\nu
\left( \nu ^{2}-4\widetilde{I}^{2}(y)\right) ^{-1/2}$.

Likewise, the asymptotically exact expression for the low frequency
conductivity, obtained in \cite{Ha-Jo:91}, coincides with our formula (\ref%
{siI}), and the correlator $C_{2}$ has the form (\ref{C2I}), after the
replacement $E_{F}\rightarrow 4E_{F}$ under the $\log $ sign. The correlator
$C_{2}$ was not considered in \cite{Ha-Jo:91}, however it can be found by
using the techniques, developed in the paper.

We note a certain difference of these asymptotically exact results
and our results. Namely, the role of the resonant distance $r(\nu )$ of (\ref%
{rnu}) in the results of \cite{Ha-Jo:91} plays (\ref{rnus}) that differs
from (\ref{rnu}) by the factor 4 under the logarithm. A possible simple
reason for this difference can be the fact that our estimate (\ref{Io}) for
the amplitude $I_{0}$ of the overlap integral indicates only its order of
magnitude, but not its precise value, or, more generally, that the
projection method is not precise enough. %As another reason we mention
%the possibility that the white noise potential can not be presented in the
%form (\ref{Veff1}) with sufficient precision. If the factor 4 is because of
%this reason, or a similar one, not related to approximations, made in the
%analysis of the two-well problem, then we have to conclude that our method
%can be applied to predict a functional form of physical quantities, but not
%to find their exact form, including numerical factors.

\medskip \noindent \textit{5.3. Weak localization regime in one dimension.}

\smallskip \noindent The case of the Gaussian white noise (\ref{wn}) in one
dimension has also been studied in the weak localization regime of large
positive energies
\begin{equation}
D^{2/3}\ll E  \label{wnwlr}
\end{equation}%
(see the works \cite{Be:73,Ab-Ry:75,LGP,Go-Co,Go}). The density of states in
this case is the free one $\rho _{0}(E)=(2\pi E)^{-1/2}$, and the
localization radius is
\begin{equation}
r_{l}=\frac{4E}{D}.  \label{rlw}
\end{equation}%
The rate of the exponential decay of wave functions is $1/r_{l}$ with
probability 1, as it was in the strong localization regime (see (\ref{Lyap}%
)).

There are several techniques that can be used in this case \cite%
{Be:73,Ab-Ry:75,LGP,Go-Co}, and yield the low frequency conductivity and the
correlators $C_{1}$ and $C_{2}$ in quadratures. It turns out that these
quantities have qualitatively the same spatial behavior as in the strong
localization regime, provided that $2E_{F}$ (i.e., $2I_{0}$ according to (%
\ref{Io})) in (\ref{si1}) is replaced by $(D/2E_{F}^{1/2})$.  Note that $%
(D/2E_{F}^{1/2})^{-1}$ coincides with the relaxation time $\tau ,$
well known from the kinetic theory \cite{Go}. According to
\cite{Ef}, the quantities $I_{0}$ and $\tau ^{-1}$ have the same
meaning: they give the order of magnitude of the difference of the
energies (spacing) of two localized states, whose centers are
separated by a distance of the order of the localization radius.
Similarly, the role of the resonant distance in the two-point
correlators $C_{1,2}$ plays (cf (\ref{rnu}) and (\ref{rnus})):
\begin{equation}
\widehat{r}(\nu )=r_{l}\log 8/\nu \tau ,  \label{rnuw}
\end{equation}%
and the rate of the exponential decay of the two-point correlators $C_{1,2}$
near the origin is $1/2r_{l}$. This rate is 4 times less than the rate $%
2/r_{l}$ of these correlators in the strong localization regime, found in
Section 3 from the naive prediction, based on the spatial behavior of the
envelope of the eigenfunctions with probability 1 (see (\ref{Lyap})), and in
\cite{Ha-Jo:91} from an asymptotically exact analysis of the corresponding
correlators. This difference can be related to the fact that eigenfunctions
in the one dimensional case in the weak localization regime are much more
spread out than in the strong localization regime. Hence, their behavior on
almost all realizations can differ from the behavior of their moments,
entering exact formulas (\ref{Ku1}), (\ref{C1}), and (\ref{C2}).

We stress that the basic properties of the strong localization regime and,
in particular, those, motivated assumptions and techniques of this paper,
are different in several important points from the basic properties of the
weak localization regime in dimension 1, where the mechanism of localization
is not trapping in deep and rare localization wells but the enhanced
backscattering due to the destructive interference between incident and
reflected waves from many defects. One of manifestations of this complex
statistical structure of wave functions in the weak localization regime is
the value of the rate of exponential decay of the correlators $C_{1,2}$,
discussed above. %In this situation the coincidence, up to a rescaling, of
%the asymptotic forms of the low frequency conductivity and the two-point
%correlators in these two extreme cases of the one-dimensional localization,
%is, in our opinion, a manifestation of a certain universality of the
%localization phenomena, the tunnelling process in particular, at least in the
%one dimensional case.
Moreover, according to \cite{Go-Co}, the characteristic length scale of the
correlators $C_{1,2}$ in the neighborhood of $r(\nu )$ is $\sqrt{r(\nu )r_{l}%
}$, i.e., is much bigger than the scale $r_{l}$ in the neighborhood of the
origin, while, according to our formulas and respective formulas of \cite%
{Ha-Jo:91}, in the strong localization regime this scale is $r_{l}$ both
near $r(\nu )$ and the origin.

\medskip \noindent \textit{5.4. Instanton approach.}

\smallskip \noindent This is a version of the variational method, proposed
first by I. Lifshitz \cite{Li:65} to find the asymptotic form of the density
of states and other characteristics of disordered systems in the strong
localization regime (see e.g. \cite{Ef-Sh:84}). The instanton approach was
used to analyze the correlators $C_{1,2}$ and the low frequency conductivity
for the white noise random potential in $d$ dimensions in paper \cite{We-Co}%
, in which the reader can find references on earlier applications
of the approach. It is based on the assumption that in the strong
localization regime the two-point correlators correspond to the
two-well potential that minimizes the total probability
distribution of the random potential under the constraints that
$H(V)\psi _{k}=E_{k}\psi _{k},\;k=1,2$ and that the well centers
of the \textquotedblright optimal\textquotedblright\ potential are
a distance $y=x_{1}-x_{2}$ apart. This has to be compared with the
DOS computation, where it is assumed that the optimal potential is
a well for which $H(V)\psi =E\psi $ (see \cite{LGP,Ef-Sh:84}). The
derivation of final formulas in \cite{We-Co} is rather involved
because of existence of two energy scales and of collective modes,
in particular those that correspond to the center of mass
$(x_{1}+x_{2})/2$ of the optimal potential (it is an analogue of
our parameter $a$ in (\ref{Cal}) - (\ref{phi2})). As a result, it
is shown in \cite{We-Co} that in the strong localization regime
(called the hydrodynamic regime in \cite{We-Co}) the correlator
$C_{1}$ and the low frequency conductivity have qualitatively the
same form as those found in Sections 3--4.

We note also that the 1-dimensional results for the white noise potential of
\cite{Ha-Jo:91} can be viewed as a justification of the instanton approach
in the one-dimensional case, because it was shown in this paper that the
two-well potential of a special from is indeed a saddle point of the
respective functional integral.

\medskip \noindent \textit{5.5. Maryland model}.

\smallskip \noindent The most widely known signature of localization is the
exponential decay of the localized states at infinity. However, the initial
derivation of the Mott formula (\ref{Mott}) as well as the above derivation
are based not only on the exponential localization, reflected in the
exponential decay of \textquotedblright bare\textquotedblright\ states of
the independent quantization in each localization well, but also on the weak
correlation between the spectra of independent quantization, reflected in
statistical independence of  localization wells in our effective potential (%
\ref{Veff}) and in appearance of the "two-hump" states in our calculations
of Sections 3, and 4. The relevance of the last property becomes clearer if
one recalls the results, obtained for an explicitly soluble model of an
incommensurate system, known as the Maryland model \cite%
{Pr-Co:82,Pa-Fi:84,Si}. This is a multi-dimensional tight binding model with
an arbitrary short-range and translation invariant hopping and with the
potential of the form
\begin{equation}
V(x)=g\tan \pi (\alpha \cdot x+\omega ),\;x\in \mathbf{Z}^{d},  \label{potM}
\end{equation}%
where $g>0$ is the coupling constant, $\alpha $ is a $d$-dimensional vector
with incommensurate components, and $\omega \in \lbrack 0,1)$ is a phase,
that plays the role of a randomizing parameter. It was found in the
mentioned papers that if for some $C>0$ and $\beta >d$ the vector $\alpha $
satisfies the Diophantine condition
\begin{equation}
|\alpha \cdot x+m|\geq C/|x|^{\beta }  \label{Dio}
\end{equation}%
for any integer $m$ and $x\neq 0$, then all the states of the model are
exponentially localized for any coupling constant, any energy, and arbitrary
dimensionality $d$ of the lattice $\mathbf{Z}^{d}$. Since the potential has
arbitrary high peaks, the model can be viewed as an explicitly soluble model
of the strong localization regime. The spectrum of the model consists of the
solutions of the equation
\begin{equation}
N(E_{t}(\omega ))=\alpha \cdot t+\omega \;(\mathrm{mod}\;1),  \label{eiv}
\end{equation}%
where $t$ is a lattice point, $N(E)=\int_{-\infty }^{E}\rho (E^{\prime
})dE^{\prime }$,
\begin{equation*}
\rho (E)=\frac{1}{\pi }\int_{\mathbf{T}^{d}}\frac{g}{\left( w(k)-E\right)
^{2}+g^{2}}dk,
\end{equation*}%
is the density of states, in which $w(k)$ is the Fourier transform of the
hopping coefficient, and $\mathbf{T}^{d}$ is the $d$-dimensional torus.

It is easy to show that for each point $t$ of the $d$-dimensional lattice
the equation has a unique solution, that if $E_{t_{1}}(\omega
)=E_{t_{2}}(\omega )$, then $t_{1}=t_{2}$, and that the set $\{E_{t}(\omega
)\}_{t\in \mathbf{Z}^{d}}$ of eigenvalues is dense for any $\omega \in
\lbrack 0,1)$.

The corresponding eigenfunctions $\psi _{t},\;t\in \mathbf{Z}^{d}$ have the
form
\begin{equation}
\psi _{t}(x)=\chi (x-t,E_{t}(\omega )),  \label{eif}
\end{equation}%
where $\chi (x,E)$ decays exponentially in $x$:
\begin{equation}
|\chi (x,E)|\leq Ce^{-|x|/r_{l}}  \label{chi}
\end{equation}%
with some positive $r_{l}(E)$. Formulas (\ref{eiv})--(\ref{chi}) seem fairly
natural in the case of the strongly incommensurate potential (\ref{potM}),
where due to the absence of any symmetry the only good quantum number to
label levels and states is the "center" of  localization well.

One can also say that Mott's notion of the localization centers is explicit
here, because, according to (\ref{eiv}) and (\ref{eif}), for any lattice
point $t$ there exists a unique eigenvalue $E_{t}$, whose eigenfunction is
exponentially localized in a neighborhood of $t$. Thus the set of
localization centers coincides with the whole lattice and the density of
localization centers, whose states have energies in a neighborhood of a
given $E$ is the density of states $\rho (E)$. This fact can be interpreted
as the uniform distribution in space of the localization centers,
corresponding to energy $E$, and is in qualitative agreement with our
assumptions of Section 2, formula (\ref{rhomu}) in particular.

On the other hand, the low frequency conductivity and the correlators $C_{1}$
and $C_{2}$ for the potential (\ref{potM}) have a rather different structure
than in the case of random potential discussed in Sections 3 and 4. This can
be seen from the form of the kernel $\rho _{E}(x,y)$, following from (\ref%
{eiv}) - (\ref{eif}):
\begin{equation}
\rho _{E}(x,y)=\sum_{t\in \mathbf{Z}^{d}}\delta (E-E_{t}(\omega ))\chi
(x-t,E_{t}(\omega ))\chi (y-t,E_{t}(\omega )).  \label{rhoM}
\end{equation}%
Consider first the correlator $C_{2}$. Plugging (\ref{rhoM}) into (\ref{C2}%
), and recalling that the averaging operation $\left\langle ...\right\rangle
$ here is the integration with respect to the parameter $\omega \in \lbrack
0,1)$ of (\ref{potM}), we find first of all that the correlator $C_{2}$ is
not a regular function. Rather, there exists a dense set of special
frequencies for which $C_{2}$ has $\delta $-peaks. If, however, we are
interested in the gross features of $C_{2}$, then we can apply a certain
smoothing procedure, say $\nu ^{-1}\int_{0}^{\nu }...d\nu ^{\prime }$. Then
we obtain that there exists the length scale
\begin{equation}
r_{1}(\nu )=\left( \frac{\nu _{0}(E)}{\nu }\right) ^{1/\beta },\;\;\nu
_{0}(E)=C/\rho (E),  \label{rMar}
\end{equation}%
(here $\beta $ and $C$ are defined in (\ref{Dio})), such that $C_{2}(x;\nu
,E)$ is of the order $e^{-r_{1}(\nu )/r_{l}}$ if $|x|\ll r_{1}(\nu )$, and $%
C_{2}(x;\nu ,E)$ is $\rho ^{2}(E)$ if $|x|>>r_{1}(\nu )$, and the transition
from the first value to the second one is in the layer $|x-r_{1}(\nu
)|\backsimeq r_{l}$, where $r_{l}$ is defined in (\ref{chi}). We see that
the qualitative form of the correlator $C_{2}$ for $|x|>>r_{l}$ is similar
to that in the random case, however there is no peak at the origin and the
length scale (\ref{rMar}) is polynomial in $\nu $ (cf (\ref{rnu})). In
addition, the length scale (\ref{rMar}) has a different origin than (\ref%
{rnu}): it is not due to the tunnelling for \textquotedblright
soft\textquotedblright\ resonant pairs, but due to the Diophantine
condition (\ref{Dio}), which determines now the distance to the
nearest localization well of an almost same energy. At low
frequencies $r_{1}(\nu )$ is much bigger than the resonance
tunnelling distance $r(\nu )$ of (\ref{rnu}). This leads to the
qualitative change of the form of the correlator $C_{1}$.
Indeed, by using the same argument, we find that $C_{1}$ is of the order $%
e^{-2r_{1}(\nu )/r_{l}}\ll 1$ for all $x$. This and formula (\ref{siC1})
imply that the low frequency conductivity is of a similar order \cite%
{Pa-Fi:84}
\begin{equation}
\sigma (\nu ,E_{F})\backsimeq \exp \{-(\nu _{1}(E)/\nu )^{1/\beta }\},\;\nu
_{1}=\frac{2^{\beta }\nu _{0}}{r_{l}^{\beta }}.  \label{PF}
\end{equation}%
The significant difference between (\ref{PF}) and (\ref{Mott}) can
be related to the absence of long range tunnelling in the Maryland
model. The spectrum of the model is too \textquotedblright
rigid\textquotedblright , the energy levels are too regularly
distributed and small level spacing are too rare for the
long-range tunnelling to happen. This illustrate the role of
resonance tunnelling in obtaining the Mott formula as well as the
range of applicability of the approach of this paper, based on
ansatz (\ref{Veff}) and on the density expansion. Besides, we see
that the low frequency conductivity provides a physical
distinction between the strong localization regimes of a random
shortly correlated and smoothly distributed potential, and the
incommensurate potential (\ref{potM}) (recall that the density of
states and the Lyapunov exponent coincide for the Maryland model
and for the random model in which the potential is a collection of
independent identically distributed Cauchy random variables, and
in which we expect our approach to be applicable). Besides,
recalling the structure of the localized states for smooth
incommensurate potentials of large amplitude in one dimension
\cite{Sin}, e.g. the potential $g\cos 2\pi (\alpha x+\omega
),\;g>>1$, one may expect that these potentials will be closer to
random potentials in the spatial behavior of two-point correlators
and the low frequency asymptotics of the conductivity.

\renewcommand{\thesection}{\Alph{section}} \setcounter{section}{1} %
\setcounter{equation}{0}

\section*{Appendix}

\subsection*{One-dimensional case with delta potentials}

To support the usage of the projection method by which the bound states of
the two-well Hamiltonian in Section 3.2 were found, we will consider here
the one-dimensional case with two delta-wells. The corresponding Hamiltonian
is:
\begin{equation}
H^{(2)}=-\frac{d^{2}}{dx^{2}}-2\sqrt{g_{1}}\delta (x-x_{1})-2\sqrt{g_{2}}%
\delta (x-x_{2}),  \label{H21}
\end{equation}%
where $g_{1,2}>0$. In this case each of two one-well Hamiltonians
\begin{equation*}
-\frac{d^{2}}{dx^{2}}-2\sqrt{g_{1,2}}\delta (x-x_{1,2})
\end{equation*}%
has the unique bound state
\begin{equation}
\varphi _{1,2}(x)=g_{1,2}^{1/4}\varphi (\sqrt{g_{1,2}}x),\;\;\varphi
(x)=e^{-|x|},  \label{psi1}
\end{equation}%
corresponding to the energy
\begin{equation}
E_{1,2}^{(1)}=-g_{1,2}.  \label{e1}
\end{equation}%
Since the Hamiltonian $H^{(2)}$ is invariant under translation, we can
replace $x_{1}$ by $0$, and $x_{2}$ by $y$. It is easy to see that $H^{(2)}$
has two bound states:
\begin{equation}
\psi _{1,2}(x)=\left[ \sqrt{g_{1}}\psi (0)\exp \left( -\sqrt{|E|}|x|\right) +%
\sqrt{g_{2}}\psi (y)\exp \left( -\sqrt{|E|}|x-y|\right) \right] \Big|%
_{E=E_{1,2}},  \label{psi21}
\end{equation}%
where $E_{1,2}^{(2)}$ are the corresponding energies. They solve the
equation:
\begin{equation}
(\sqrt{|E|}-\sqrt{g_{1}})(\sqrt{|E|}-\sqrt{g_{2}})=\sqrt{g_{1}g_{2}}\exp
\left( -2\sqrt{|E|}|y|\right)   \label{E21}
\end{equation}%
Assuming the same accuracy as in Section 3.2 ($|g_{1}-g_{2}|\ll
g_{1,2},\;|y|>>g_{1,2}$), we find that the solutions $E_{1,2}^{(2)}$ of (\ref%
{E21}) have the form (\ref{levs}) in which $I(y)=2ge^{-\sqrt{g}|y|}$ (cf (%
\ref{Ias}) - (\ref{Io})), and the eigenfunctions (\ref{psi1}) have the form (%
\ref{psi}) - (\ref{tan}) in which $\varphi _{1,2}$ are given by (\ref{psi1}).

Another way to act in this case is to plug the exact states and levels,
given by (\ref{psi21}) - (\ref{E21}), into the expressions (\ref{si2}), (\ref%
{C12}) and (\ref{C220}) for the two well contributions for the conductivity
and the correlators $C_{1}$, and $C_{2}$. This leads to rather complicated
formulas which, however, have the same asymptotic behavior as our formulas (%
\ref{siI}), (\ref{C1I}), and (\ref{C2I}) in the asymptotic regime $0<\nu \ll
|E|$.

\end{document}